\documentclass[aps,prd,tighten,11pt,showpacs,a4,nofootinbib]{revtex4}
\usepackage[utf8]{inputenc}

\usepackage{graphicx}
\usepackage{bm}
\usepackage{amsmath,amssymb}
\usepackage{latexsym}

\newcommand{\bee}{\begin{equation}}
\newcommand{\eee}{\end{equation}}

\begin{document}

\title{Quasinormal modes of black holes. The improved semianalytic approach.}

\author{Jerzy Matyjasek}\email{jurek@kft.umcs.lublin.pl} 
\author{Micha{\l} Opala}

\affiliation{Institute of Physics,
Maria Curie-Sk\l odowska University\\
pl. Marii Curie-Sk\l odowskiej 1,
20-031 Lublin, Poland}

\begin{abstract}
We have extended the semianalytic technique of Iyer and Will for computing 
the complex quasinormal frequencies of black holes, $\omega,$ by constructing the Pad\'e approximants
of the (formal) series for $\omega^{2}$. It is shown that for the (so far best documented)
quasinormal frequencies of the Schwarzschild and Reissner-Nordstr\"om black holes 
the Pad\'e transforms $P_{6}^{6}$ and $P_{7}^{6}$ are, within the domain of applicability,  
always in excellent agreement with the 
numerical results. We argue that the method may serve as the black box with the 
``potential'' $Q(x)$  as an input and 
the accurate quasinormal modes as the output. The generalizations and modifications of the method are 
briefly discussed as well as the preliminary results for other classes of the black holes.
\end{abstract}
\maketitle


\section{Introduction}

The physical black holes are not isolated systems, they interact in a variety of ways with their environment
and changing their surrounding they change themselves. Especially interesting in this regard  are the perturbations
that can satisfactorily be described within the linear approximation. 
On general grounds one expects that the late-time
behavior is dominated by the oscillations that are characteristic to a given black hole and independent of the initial cause 
of the perturbation. That means that the gravitational wave emitted by the perturbed black hole will carry the imprints 
of its characteristics on the unique set of complex numbers, $\omega,$ simultaneously  describing the rates of damping and the 
frequency of the oscillations. Indeed, numerical analysis of the evolution of the black holes formed in a gravitational collapse or 
in the collision of black holes indicates that each of them approaches such a ringdown phase.
These quasinormal oscillations are expected to be crucial both in the black holes detection
and, when discovered, in studying their properties. It is natural that the quasinormal oscillations of the black holes 
have been area of intense study for the last 40 years.

The quasinormal modes  considered in this paper are the solutions of the second-order differential equation
\begin{equation}
 \frac{d^{2}}{dx^{2}} \psi(x) + Q(x) \psi(x) =0,
 \label{schroedinger}
\end{equation}
where  $-Q(x)$ is a potential function, which is assumed to be constant as $|x|\to \infty$ (the limits
may be different) and to posses
the maximum at some finite $x_{0},$ subjected to the particular set of the boundary conditions. The function $\psi(x)$ 
is the radial part of the free oscillations (with the assumed  time-dependence of the 
form $e^{-i \omega t}$) which is purely ``outgoing'' as $|x|\to \infty.$ Here we follow convention proposed
in Ref.~\cite{wkb0} and understand the term ``outgoing'' as ``moving away from the potential barrier''.
For a perturbation of a given spin weight, $s,$ the quasinormal modes are labeled by the multipole number, $l,$ 
and the overtone number $n.$

The quasinormal modes have been studied both numerically and analytically for the various
perturbations of the black hole backgrounds. Especially interesting are the analytic or semianalytic
methods allowing quick and accurate calculations for a wide range of black holes.
Currently there are a few popular approaches to the quasinormal frequencies problem, each having its own merits.
A high reputation  of the continued fraction method is due to its great accuracy and possibility to calculate
high overtones~\cite{Leaver1,Leaver1a}. In its original form it has been employed in the three term recurrence relation, however, 
the more complicated cases can also be addressed by reducing them to three term recurrence by Gauss elimination~\cite{Leaver2}. 
The Hill-determinant method proposed in Ref.~\cite{Pancha} is in a sense complementary to the Leaver approach
and allows for calculations of the high overtones by searching of the stable zeros of the high-order polynomials.
Nollert~\cite{Nollert1,Nollert2} studied the quasinormal modes via Laplace transform and analyzed the problem
of overtones.
The quasinormal modes
(of the Schwarzschild black hole)  have been calculated by Zaslavskii by reducing the problem to the well-known
quantum anharmonic oscillator~\cite{oleg}.
The competitive approaches include various incarnations of the phase integral method~\cite{phase1,phase2,galtsov} 
and the modifications of the WKB approximation~\cite{wkb0,wkb1}.
The Iyer-Will method~\cite{phase1}, which belongs to the latter class, and its generalization to the sixth order~\cite{roman1} gained great
and well-deserved popularity.
(See for example Refs.~\cite{Stuchlik,Fernando1,Bezerra,Fernando2,lin,Stuchlik2} and the references cited therein).

Typically, depending on the character of the problem, one is torn between 
the need for a  high accuracy  (also for overtones $n\gg l$)
and the generality of the approach, allowing analysis  of various 
potentials, even at the expense of some inherent limitations.
Practically, these limitations may not be so serious as for the astrophysical black holes  
the least damped modes are most significant and simultaneously easiest to calculate.
In our opinion the WKB method is a best choice due to its generality and flexibility and may serve, 
with necessary modifications, as the black box with the ``potential'' $Q(x)$  as an input and 
the quasinormal modes as the output.  
In this paper we shall propose a modification of the Iyer-Will method~\cite{wkb1,roman1}. 
The modification is twofold: First we generalize their approach by extending calculations beyond sixth-order 
WKB and subsequently employ the powerful technique of the Pad\'e transforms. Restricting to the so far best documented 
qnasinormal modes of the four dimensional Schwarzschild  and  Reissner-Nordstr\"om black holes
we show that the approximation works very well. Typically, the  deviations of the real and imaginary part
of the complex frequencies from the accurate numerical results are  
smaller than  those obtained within the framework of competing approaches. Moreover, for the low-lying modes the accuracy is comparable with or even better
than the phase integral method in the optimal order. 
     
The paper is organized as follows. In Sec.~\ref{mett} we shall briefly introduce the method. In Sec.~\ref{rach}
we will discuss the results obtained for the various perturbations in the Schwarzschild and Reissner-Nordstr\"om black holes
and make a detailed comparison with the accurate numerical calculations. Finally, in Sec.~\ref{last} we shall 
we shall briefly discuss the possible extensions and modifications of the method as well as our preliminary results
for other classes of black holes.

\section{The method\label{mett}} 
As is well known the modification of the WKB approach proposed by Iyer and Will~\cite{wkb1,wkbOvertones,carlB} 
allows for configurations with  closely lying classical turning points. The idea is to match 
simultaneously exterior WKB solutions across the two turning points. 
The differential equation (\ref{schroedinger}) in the interior region is first simplified by expanding 
the potential into the Taylor series up to  terms of order six
and subsequently solved (approximately) in terms of the parabolic cylinder functions.
The asymptotic approximation to the interior solution is used to match the third-order WKB solutions.
This approach has been used to calculate the approximate quasinormal frequencies of the
Schwarzschild~\cite{wkb2}, Reissner-Nordstr\"om~\cite{wkb3} and  Kerr~\cite{wkb4}  black holes. A more profound
comparison  shows that the approximate results deviate (at worst) from the accurate numerical ones by a few percents.
A natural question arises: Is it possible to modify the Iyer-Will approximation and get better results?
Below we shall show that the answer to this question is affirmative.

We start with the presentation of some basic informations concerning the strategy adopted in this paper. 
The method is largely due to Iyer and Will but with some necessary generalizations and modifications. 
Following Ref.~\cite{wkb1}
let us consider the two-turning-point problem with the turning points located at, say, $x_{1}$ and $x_{2}$ $(x_{1} < x_{2}).$ 
We can distinguish three regions: The interior region (II) between the turning points and the regions  I and III  outside
the points $x_{1}$ and $x_{2}$, respectively. In the exterior regions the (asymptotic) solution
is given by the standard WKB solution constructed to the required order. Substituting
\begin{equation}
 y(x) \sim \exp(S/\varepsilon),
\end{equation}
where 
\begin{equation}
S(x) = \sum_{k=0}^{\infty} \varepsilon^{k} S_{k}(x)
\end{equation}
to \begin{equation}
\varepsilon^{2}  \frac{d^{2}}{dx^{2}} \psi(x) + Q(x) \psi(x) =0,
 \label{schroedinger1}
\end{equation}
and collecting the terms with the like powers of $\varepsilon$ one obtains a chain of equations
of ascending complexity. 
The expansion parameter $\epsilon,$ which helps to keep the order control of the complicated terms should 
be set to 1 at the end of the calculations.
The first two solutions, $S_{0}$ and $S_{1},$ define the standard WKB 
approximation. 
On the other hand, in the region II, the strategy is different. First we expand $Q(x)$ into
the Taylor series about the maximum  of $-Q(x)$   (located at $x_{0}$) to the required order
and making use of the substitutions
\begin{equation}
 k = \frac{1}{2} Q''_{0}, \hspace{5mm} z_{0}^{2} = -2 Q_{0}/ Q''_{0},
\end{equation}
and
\begin{equation}
 s_{n} = \frac{2 Q^{(n)}_{0}}{n!  Q''_{0}}, \hspace{5mm} (n=3,...)
\end{equation}
we rewrite Eq.~(\ref{schroedinger1}) in the following form
\begin{equation}
  \varepsilon^{2}  \frac{d^{2}}{dz^{2}} \psi + k ( -z_{0}^{2}+ z^{2} + \sum_{n=3}^{2 N} s_{n} z^{n}) \psi =0,
\end{equation}
where $z=x-x_{0}.$ With the aid of
\begin{equation}
 t =(4k)^{1/4} e^{-i \pi/4} z/\varepsilon^{1/2},
\end{equation}
\begin{equation}
 \bar{s}_{n} = \frac{1}{4} s_{n} (4 k)^{(2-n)/4} e^{n i \pi/4} 
\end{equation}
and
\begin{equation}
 \nu+\frac{1}{2} = -i k^{1/2} z_{0}^{2}/2 \varepsilon - \sum_{n=2}^{N} \varepsilon^{n-1} \Lambda_{n}
 \label{nu_plus}
\end{equation}
this equation can be further transformed 
\begin{equation}
 \frac{d^{2}\psi}{dt^{2}} + \left\{\nu+\frac{1}{2} -\frac{1}{4} t^{2} -\sum_{n=1}^{N-1} \left[\varepsilon^{(2n-1)/2} \bar{s}_{2n+1}t^{2n+1}
 + \varepsilon^{n}\left( \bar{s}_{2n+2} t^{2n+2} - \Lambda_{n+1}\right) \right]  \right\} \psi =0.
 \label{cylinder}
\end{equation}
In the limit $\varepsilon \to 0$ the general solution of the equation  (\ref{cylinder}) is the linear combination
of the parabolic cylinder functions
\begin{equation}
 a D_{\nu}(t) + b D_{-\nu-1}(-it).
\end{equation}
In a general case we will look for a solution of the form
\begin{equation}
 \psi(t) = f(t) D_{\nu}(g(t)),
\label{parabol2}
 \end{equation}
where $f(t)$ and $g(t)$ are two functions that have to be determined. Substituting (\ref{parabol2}) into (\ref{cylinder})
and eliminating the term with the first derivative of the parabolic cylinder function one  gets $f=-(dg/dt)^{-1/2}.$
As a result one obtains the differential equation for the function $g(t)$ which can be solved assuming
\begin{equation}
 g(t) = t + \sum_{n=0}^{\infty} \varepsilon^{n/2} A_{n}(t),
\end{equation}
where 
\begin{equation}
 A_{n}(t) = \sum_{j=0} \alpha_{2j}^{n} t^{n+1-2j}, \hspace{5mm} (n+1-2j\geq0).
\end{equation}
It should be noted that by solving the resulting system of algebraic equations one obtains both the coefficients $\alpha$ and $\Lambda.$
The whole procedure is algorithmic and can easily be implemented in any of the existing computer algebra system.
Now, if we substitute $\Lambda_{i}$ into (\ref{nu_plus})
we will obtain the equation that relates $\omega,$
derivatives (up to $2N$) of the potential at $x_{0}$ and $\nu.$ 

Thus far the analysis has been carried out in the 
interior region. In the exterior regions the solutions are constructed within the framework of  $N$th WKB method
starting with physical optics approximation as the lowest order approximation. The asymptotic matching is 
to be performed outside the turning points, where $|t| \to \infty.$ 
The two different asymptotic representation of the parabolic cylinder function valid in a wedge $|\arg t| <3/4$
and  $1/4 \pi< \arg t < 5/4 \pi,$ respectively, which are used to represent the function $\psi$
\begin{equation}
 \psi \sim \left(\frac{dg}{dt}\right)^{-1/2} \left[ A D_{\nu}(g) + B D_{-\nu-1}(i g)\right],
\end{equation}
where $A$ and $B$ are constants,
are  matched with the ``right'' and ``left'' WKB approximants. 
The analysis of the structure of the asymptotic form of the parabolic cylinder function in the 
wedge $1/4 \pi< \arg t < 5/4 \pi$ for large $|t|$ shows that it  contains the functions $\Gamma(-\nu)$
and $\Gamma(\nu+1)$ in the denominator. In the regions I and III one has linear combination of
the appropriate WKB solutions representing incoming and outgoing waves. Hence the problem reduces to
the construction of the formulas that relate amplitudes of the waves in the asymptotic regions of I and III.
The exact matching is extremely tedious and has been carried out up to the third order WKB approximation.
However, as has been argued by Iyer and Will, the matching coefficients should depend only on $\nu,$
and, consequently, one can consider the interior problem only. Finally, imposing the black hole quasinormal mode
conditions which state that there should be no incoming modes in the region I and III, one concludes that 
$\nu$ is a nonnegative integer. Consequently the formula (\ref{nu_plus}) relating $\omega,$  derivatives of $Q$ at the $x=x_{0}$
and $\nu =n$ $(n=0,1,2,...,)$ is 
\begin{equation}
 \frac{ i Q_{0}}{\sqrt{  2Q''_{0}}} -\sum_{k=2}^{N} \Lambda_{k}= n+ \frac{1}{2},
 \label{nuplus}
\end{equation}
where each $\Lambda_{k}$ is a function of the derivatives of $Q(x)$ of ascending complexity.
The first two terms $\Lambda_{2}$ and $\Lambda_{3}$ has been constructed in Ref~\cite{wkb1}.
This result has been extended to the sixth order WKB by Konoplya in Ref.~\cite{roman1}.
It should be noted that the third order result has been reconstructed using the phase integral~\cite{froeman}
by Gal'tsov and Matiukhin in a very interesting development in Ref.~\cite{galtsov}.

In Refs.~\cite{wkb1,roman1} the terms $\Lambda_{k}$ in Eq.~(\ref{nuplus}) were summed. On the 
other hand we do not know in advance if  adding the next term will improve or worsen the quality of the approximation.
Similarly, we do not know if there is some optimal truncation. Consequently, it may be that summation of
the $\Lambda_{k}$ terms is not the best strategy.

In this paper we  report on the extension of  the results of Iyer and Will and of Konoplya 
to the 13th order WKB approximation, i.e., in addition to  $\Lambda_{2}$ and $\Lambda_{3}$  
calculated in Ref.~\cite{wkb1} and $\Lambda_{4},\, \Lambda_{5}$ and $\Lambda_{6}$
presented in Ref.~\cite{roman1} (in a few cases the typographical errors have been detected) 
we have calculated all $\Lambda_{n}$ up to $n=13.$
The novelty of our method  consists in construction, for any given  $l$ and $n,$ 
the Pad\'e approximants. As have been observed by Bender and Orszag~\cite{carlB}
  ``... Pad\'e approximants often work quite well, even beyond their proven range of
applicability...''
and we believe that making use of this two powerful techniques will improve the quality of the results.
In the next section we shall explicitly demonstrate that this is indeed the case.

The technique of the Pad\'e approximants has been used in Ref.~\cite{wkb4}
in the calculations of  the gravitational perturbations of the Kerr black hole and 
in Ref.~\cite{wkbOvertones}  but in a different 
context. Let us assume, for simplicity, that the function $Q(x)$ is of the form
\begin{equation}
 Q(x) = \omega^{2} -V(x),
 \label{narrow}
\end{equation}
where $V(x)$ does not contain $\omega.$
Now,  we approximate the right hand side of the equation
\begin{equation}
 \omega^{2} = V(x)-i \left(n+\frac{1}{2} \right)  \sqrt{  2Q''_{0}} \varepsilon -i \sqrt{  2Q''_{0}} \sum_{i=2}^{N} \varepsilon^{j} \Lambda_{j}
\end{equation}
(treated as a polynomial of $\varepsilon$) by $P_{6}^{6}$ and $P_{7}^{6}.$
Note that the  functions $\Lambda_{k}$ depend on $n.$
We have chosen to work with $P_{6}^{6}$ and $P_{7}^{6}$ although all table 
of approximants $P_{\tilde{m}}^{\tilde{n}}$ satisfying $\tilde{m} +\tilde{n} +1 \leq 14$
can be constructed.
We have not attempted to analyze the optimal (from a point of view of calculational effectiveness)  approximants and prefer to
work with $\omega^{2}.$ 

As is well known the methods of calculating frequencies of the  
quasinormal modes based on the WKB approximation break  down when the overtone number $n$
exceeds the angular harmonic index, $l,$  so one expects the reasonable results only for 
$ n\leq l,$ or $n$ slightly bigger than $l.$   On the other hand the results are 
progressively better with increasing $l.$ 

The formulas describing the higher-order $\Lambda_{n}$ are rather long and 
complicated and the calculation is time-consuming~\footnote{The Iyer-Will 
formulas can be reproduced in a few second on a typical budget laptop. On the
other hand  calculations of the general terms up to $\Lambda_{13}$ can take a few 
hours}. Indeed, the number of terms and their complexity quickly grows with 
$n$ as can be seen in Tab~\ref{tabb1}.
Fortunately the calculations of the general terms  have to be executed 
only once, and to avoid proliferation of extremely long formulas we do 
not present them here.  All $\Lambda_{n}$ $(n=2,...,13)$ stored in various 
formats\footnote{The results are stored in  Mathematica, Maple and Maxima syntax.} can be obtained from the first author 
upon request. On the other hand, the calculation of the quasinormal modes
for a given potential function is quite fast.

\begingroup\squeezetable
\begin{table}
 \caption{\label{tabb1} The number of terms in $\Lambda_{k}$ for $ 2 \leq k \leq 13. $}
\begin{ruledtabular}
\begin{tabular}{ccccccccccccc}
$\Lambda_{n}$ & $\Lambda_{2}$ & $\Lambda_{3}$ & $\Lambda_{4}$ & $\Lambda_{5}$ & $\Lambda_{6}$ & 
$\Lambda_{7}$ & $\Lambda_{8}$ & $\Lambda_{9}$ & $\Lambda_{10}$ & $\Lambda_{11}$ & $\Lambda_{12}$ & $\Lambda_{13}$ \\ \hline
Number of terms & 6 & 20 &  55 & 132 & 294 & 616 & 1215 & 2310 & 4235 & 7524 & 13026& 22050 \\ 
\end{tabular}
 \end{ruledtabular}
 \end{table}
 \endgroup
 
 Although the $\Lambda_{j}$ are complicated products of the various powers of the derivatives of the function $Q$ 
 evaluated at the maximum they can easily be calculated numerically. 
 Indeed, for each $j$ the maximal  derivative of $Q$ is $2j$ and the length of each term in $\Lambda_{j}$ is $L^{(2j-1)/2}$
 (where $Q^{(k)}$ - the $k$th derivative of $Q$ has the length $L^{-k}$) and the calculations reduce to simple multiplications.
 The derivatives of Q with respect to the general tortoise coordinate are easily programmable and the determination of the 
 radial coordinate of the maximum of the potential requires only elementary numerics.
 
 \section{The normal modes of  black holes\label{rach}}
 \subsection{Schwarzschild black hole}
  Here we give only basic informations necessary to calculate the quasinormal modes.
The odd-parity perturbation of the Schwarzschild black hole are governed by the equation
\begin{equation}
 \frac{d^{2} \psi}{dx^{2}_{\star}} + \left[\omega^{2}-\left(1-\frac{2}{x}\right) 
 \left(\frac{\lambda_{1}}{x^{2}} +\frac{2 \beta}{x^{3}} \right)\right] \psi,
\end{equation}
where $x_{\star}$ is the Regge-Wheeler coordinate, $\lambda_{1} = l(l+1)$ and $\beta = 1, 0, -3$ for the scalar, vector  
and gravitational perturbations, respectively. For the gravitational perturbations we shall also calculate the  quasinormal modes for the 
even-parity (Zerilli) potential
\begin{equation}
  \frac{d^{2} \psi}{dx^{2}_{\star}} + \omega^{2} \psi -\frac{2 \Delta}{x^{5}(\lambda_{2} x +3 )^{2}}
  \left[\lambda_{2}^{2}(\lambda_{2}+1) x^{2} + 3 \lambda_{2} x +9  \right] \psi,
\end{equation}
where $\Delta= x^2- 2x$ and $\lambda_{2} = (l-1)(l+2)/2.$
The normal mode frequencies should be the same for the both potentials and this can be used for the consistency check.

Now we shall analyze the accuracy of the WKB method with the Pad\'e transform and contrast the thus obtained results
with the numerical calculations of Fr\"oman et al. presented in Ref.~\cite{phase1}. The frequencies of the odd-parity modes 
of the scalar, electromagnetic and gravitational are given in Tables~\ref{tabb2},~\ref{tabb3} and \ref{tabb4}, respectively.
The results of the calculations of the gravitational even parity modes are presented in Table~\ref{tabb5}. 
Although it is quite  possible that the order of the Pad\'e approximants is not always optimal,
we have decided to focus our attention only on $P_{6}^{6}$ and $P_{7}^{6}$.
Such a detailed comparison is possible only if the results of the numerical calculations are known in advance. 
Typically, for a new $Q(x),$ such results are absent and we have to choose 
the calculational strategy (to certain extent) blindly.\footnote{Our codes 
are written in such a way that one can calculate any element of the table of the Pad\'e approximants.} 
The Pad\'e and numerical results are compared with the analogous result calculated within the framework of the  very
popular sixth-order ``pure'' WKB method. The appropriate formulas have been constructed previously by Konoplya. 
Our results have been obtained as a by-product of the calculations of the $\Lambda_{13},$  and, when applied to
the perturbations of the Schwarzschild black hole they extend the calculations of Ref.~\cite{roman1}.

The general features of the approximation based on the WKB method are shared by Pad\'e approximants: the quality of
the approximation increases with the increase of the harmonic index, $l,$ and deteriorates with the the overtone number $n.$
The best results are obtained for $n\leq l$ though they remain reasonable if the overtone number slightly excesses $l.$
On the other hand, the quality of the approximation within the region of validity is really superb and the quasinormal frequencies
are close to the accurate numerical results. Generally, as expected, they are also more accurate  than those calculated using sixth-order 
WKB. Moreover, the Pad\'e approximants $P_{6}^{6}$ and $P_{7}^{6}$  give $\omega$ that are comparable with or even better than 
the phase-integral method in the optimal order. Of course, in order to construct better approximants one has to take into
account the influence of the remaining turning points.

To analyze the quality of the Pad\'e approximants in more details let us define deviation of the real part
of the frequency
\begin{equation}
 \Delta^{(r)}(\omega_{k}) = \frac{\Re(\omega_{k})-\Re(\omega_{num})}{\Re(\omega_{num})} 100\%
\end{equation}
and similarly the deviation of its imaginary part
\begin{equation}
 \Delta^{(i)}(\omega_{k}) = \frac{\Im(\omega_{k})-\Im(\omega_{num})}{\Im(\omega_{num})} 100 \%,
\end{equation}
where $\omega_{k}$ is the approximate complex frequency of the quasinormal mode and $\omega_{num}$
is its accurate numerical value. 
First, let us consider the odd-parity scalar modes.
Inspection of Table~\ref{tabb2} shows that both $P_{6}^{6}$ and
$P_{7}^{6}$ are very accurate and except for the lowest mode $(l=0, n=0)$ it is always superior to
the sixth-order WKB. Indeed, $\Delta^{(r)}$ calculated for $P_{6}^{6}$ and $P_{7}^{6}$ approximants of the lowest
mode is $7.9 \times 10^{-1}\%$ and $9.8 \times 10^{-2}\%,$ respectively, whereas within the framework 
of WKB approximation one has $1.2 \times 10^{-2}\%.$ (The deviation of $P^{5}_{6}$ is slightly smaller).
On the other hand, deviation of the imaginary part of  $P_{6}^{6}$ and $P_{7}^{6}$ for that mode
is $-8.6\times 10^{-3}\%$ and $2.8\times 10^{-2}\%,$ respectively. This may be compared with deviation
calculated within the framework of the sixth-order WKB, which is $3.9\%.$
For the lowest modes $(l \geq 1)$ 
and overtones the Pad\'e approximants  are at least two order of magnitude better that the WKB. (For $l=3,n=0$
both $P_{6}^{6}$ and $P_{7}^{6}$  are exactly the same as the numerically calculated frequencies).
The imaginary part of the Pad\'e approximants to the  quasinormal modes is always far better than those calculated
using WKB.

The Pad\'e approximants for the vector and gravitational odd-parity modes are amazingly accurate as 
can be seen from Table~\ref{tabb3} and~\ref{tabb4}. Now, let us analyze the gravitational even-parity modes.
The results of the calculations are displayed in Table~\ref{tabb5}. In Table~\ref{tabb5a} the deviations $\Delta^{(r)}$
and $\Delta^{(i)}$ are presented for all calculated modes. The accuracy of the Pad\'e approximants is excellent
and ranges from $5.2 \times 10^{-1}\%$ for $(l=2, n=3)$ to $10^{-8}\%$ for $(l=4,n=1).$ This can be contrasted with
the WKB results, which deviates by $3.9\%$ from the exact value for  $(l=2, n=3)$ and $1.6\%$ for $(l=3,n=5).$ 

\begingroup\squeezetable
\begin{table}
 \caption{\label{tabb2} The quasinormal modes of the odd-parity scalar 
 perturbations of the Schwarzschild black hole. Numerical results are 
 taken from Ref.~\cite{phase1}. The WKB results have been obtained 
 previously in~\cite{roman1}. Here we have calculated them once again 
 and retained more digits. }
\begin{ruledtabular}
\begin{tabular}{cccccc}
$ l$ & $n$ & \text{Numerical value of $\omega$} & Regge-Wheeler potential $P_{6}^{6}$  & Regge-Wheeler potential $P_{7}^{6}$ &
   {\rm Sixth order WKB}   \\ \hline
 0 & 0 & 0.1104543-0.1048943 i & 0.111328786-0.104885261 i & 0.110346431-0.104924037
   i & 0.110467018-0.100816251 i \\
 & 1 & 0.0861169-0.3480524 i & 0.0869872660-0.347994694 i &
   0.0874777581-0.347875503 i & 0.0890291001-0.344528856 i \\
 1 & 0 & 0.29293610-0.097660 i & 0.292936154-0.0976600269 i & 0.292936143-0.0976599978
   i & 0.292909644-0.0977616179 i \\
 & 1 & 0.26444865-0.30625739 i & 0.264443075-0.306258587 i &
   0.264446919-0.306257973 i & 0.264471051-0.306518241 i \\
 & 2 & 0.229539335-0.540133425 i & 0.229227121-0.540073930 i &
   0.229216213-0.540091386 i & 0.231014233-0.542165408 i \\
 & 3 & 0.203258386-0.788297823 i & 0.202925543-0.788885062 i &
   0.202747941-0.788724995 i & 0.222094287-0.795168920 i \\
 2 & 0 & 0.483643872-0.096758776 i & 0.483643872-0.0967587762 i &
   0.483643872-0.0967587759 i & 0.483641882-0.0967660891 i \\
 & 1 & 0.463850579-0.295603937 i & 0.463850543-0.295603953 i &
   0.463850565-0.295603954 i & 0.463846744-0.295626915 i \\
 & 2 & 0.430544054-0.508558402 i & 0.430543707-0.508556964 i &
   0.430544040-0.508556573 i & 0.430385788-0.508699895 i \\
 & 3 & 0.393863063-0.738096585 i & 0.393853836-0.738081384 i &
   0.393855027-0.738098967 i & 0.393206798-0.739885188 i \\
\end{tabular} 
 \end{ruledtabular}
 \end{table}
\endgroup

\begingroup\squeezetable
\begin{table}
 \caption{\label{tabb3}  The quasinormal modes of the odd-parity electromagnetic perturbations of the Schwarzschild black hole. Numerical results are 
 taken from Ref.~\cite{phase1}. The WKB results have been obtained previously in~\cite{roman1}. Here we have calculated them once again and retained more digits.}
\begin{ruledtabular}
\begin{tabular}{cccccc}
$ l$ & $n$ & \text{Numerical value of $\omega$} & Regge-Wheeler potential $P_{6}^{6}$  & Regge-Wheeler potential $P_{7}^{6}$ &
   {\rm Sixth order WKB}  \\  \hline
 1 & 0 & 0.248263272-0.092487709 i & 0.248263238-0.0924876590 i &
   0.248263273-0.0924877469 i & 0.248191418-0.0926370275 i \\
 & 1 & 0.21451542-0.293667646 i & 0.214496169-0.293664604 i &
   0.214503194-0.293668427 i & 0.214295290-0.294118148 i \\
  & 2 & 0.174773568-0.525187599 i & 0.174483574-0.524812107 i &
   0.174838137-0.525458762 i & 0.173988785-0.530058622 i \\
  & 3 & 0.146176699-0.771908924 i & 0.144481853-0.772141225 i &
   0.147368451-0.772137961 i & 0.159002152-0.796775458 i \\
 2 & 0 & 0.457595512-0.095004426 i & 0.457595511-0.0950044260 i &
   0.457595512-0.0950044257 i & 0.457593408-0.0950111516 i \\
  & 1 & 0.436542386-0.290710143 i & 0.436542317-0.290710154 i &
   0.436542359-0.290710162 i & 0.436533960-0.290728026 i \\
  & 2 & 0.401186734-0.501587346 i & 0.401189409-0.501584248 i &
   0.401185685-0.501588318 i & 0.400905984-0.501728128 i \\
  & 3 & 0.362595032-0.730198514 i & 0.362585402-0.730166653 i &
   0.362589034-0.730210508 i & 0.361158598-0.732452614 i \\
 3 & 0 & 0.656898671-0.095616218 i & 0.656898670-0.0956162179 i &
   0.656898670-0.0956162179 i & 0.656898466-0.0956171153 i \\
  & 1 & 0.641737436-0.289728402 i & 0.641737435-0.289728402 i &
   0.641737436-0.289728402 i & 0.641736559-0.289730658 i \\
  & 2 & 0.613832026-0.492066258 i & 0.613832147-0.492066236 i &
   0.613832001-0.492066269 i & 0.613787638-0.492063925 i \\
  & 3 & 0.577918506-0.70633083 i & 0.577919257-0.706329774 i &
   0.577918194-0.706330849 i & 0.577481153-0.706497836 i \\
\end{tabular} 
\end{ruledtabular}
 \end{table}
\endgroup

\subsection{Reissner-Nordstr\"om black hole}
Technically speaking, the equations governing the  perturbations of the Reissner-Nordstr\"om black hole are only slightly more complicated 
than the  analogous equations for the Schwarzschild black hole. The main difference lies in the fact that for 
the charged black holes the are no pure electromagnetic or pure gravitational oscillations.  
The (odd-parity) normal oscillations are  governed by the system of the differential equations
\begin{equation}
\frac{d^{2} \psi^{(-)}_{j}}{dx_{\star}^{2}} + \omega^{2} \psi^{(-)}_{j} - \frac{\Delta}{r^{5}} \left[ \lambda_{1} r -p_{k} + \frac{4 q^{2}}{x}\right] \psi^{(-)}_{j}, \hspace{0.5cm} (j \neq k)
\label{pott}
\end{equation}
where $q=|e|/M$ with $e$ being the electric charge,   $\Delta = x^{2} -2 x+ q^{2},$ $i,j =1,2$ and
\begin{equation}
 p_{1} = 3 +\left( 9 + 8\lambda_{2} q^{2} \right)^{1/2},
\end{equation}
\begin{equation}
 p_{2} = 3 -\left( 9 + 8\lambda_{2} q^{2} \right)^{1/2}.
\end{equation}
The tortoise coordinate $x_{*}$ is defined in the standard way.
Note that for uncharged black hole $q=0$ the functions $\psi^{(-)}_{1}$  and $\psi^{(-)}_{2}$ 
describe electromagnetic and gravitational perturbations, respectively.
The last term in Eq.~(\ref{pott}) defines the minus potential, i.e.,  $(-V^{(-)}_{j}\psi^{(-)}_{j}) .$ 
The potential of the even parity modes
is given by
\begin{equation}
 V_{j}^{(+)} = V_{j}^{(-)} + 2 p_{i} \frac{d}{dx_{\star}}\left( \frac{\Delta_{1}}{2 \lambda_{2} r^{2} + p_{i}} \right)
\end{equation}
and the perturbation equations are given  by (\ref{pott}) with the last 
term in the right hand side replaced by $-V^{(+)}_{j}\psi^{(+)}_{j} $ and $\psi^{(-)}_{j} $ by   $\psi^{(+)}_{j}. $

The quasinormal modes of the Reissner-Nordstr\"om black hole have been studied 
by a number of authors~\cite{Leaver2,AnderssonRN1,wkb3,Onozawa2,Onozawa1,Konoplya2,Hod}. 
Here we restrict ourselves to the odd-parity electromagnetic and gravitational 
modes as the extension to the even-parity modes does not present any problems and the 
results are of the same quality.
The Pad\'e approximants of the quasinormal frequencies are compared with 
the accurate numerical calculations of Andersson~\cite{AnderssonRN1} for the nonextreme black holes
and with the results of Onozawa et al.~\cite{Onozawa1} for the extreme ones.
The results are displayed in Tables~\ref{tabb6}-\ref{tabb8}. A more detailed 
comparison of the Pad\'e approximants of the gravitational modes with 
the accurate numerical results is given
in Tables~\ref{tabb9} and ~\ref{tabb10}. It is seen that the Pad\'e approximants 
are very close to the numerical $\omega$  and the accuracy of the results does not depend on $q.$
Similarly, for the extreme 
configuration the Pad\'e approximants give excellent agreement with the numerical results presented in Ref.~\cite{Onozawa1}.

\begingroup\squeezetable
\begin{table}
 \caption{\label{tabb4}   The quasinormal modes of the odd-parity gravitational 
 perturbations of the Schwarzschild black hole. Numerical results are 
 taken from Ref.~\cite{phase1}. The WKB results have been obtained previously
 in~\cite{roman1}. Here we have calculated them once again and retained more digits.}
\begin{ruledtabular}
\begin{tabular}{cccccc}
$ l$ & $n$ & \text{Numerical value of $\omega$} & Regge-Wheeler potential $P_{6}^{6}$  & Regge-Wheeler potential $P_{7}^{6}$ &
   {\rm Sixth order WKB}  \\ \hline
 2 & 0 & 0.373671684-0.088962315 i & 0.373671489-0.0889707371 i &
   0.373672999-0.0889666261 i & 0.373619357-0.0888909796 i \\
  & 1 & 0.346710997-0.273914875 i & 0.346285174-0.273927461 i &
   0.346735243-0.273772106 i & 0.346296571-0.273479774 i \\
  & 2 & 0.301053455-0.478276983 i & 0.299166032-0.478552363 i &
   0.301756472-0.476569959 i & 0.298519956-0.477560223 i \\
  & 3 & 0.251504962-0.705148202 i & 0.248821369-0.707230614 i &
   0.244063926-0.695706014 i & 0.241715477-0.709595458 i \\
 3 & 0 & 0.599443288-0.092703048 i & 0.599443289-0.0927030478 i &
   0.599443290-0.0927030479 i & 0.59944337-0.0927025232 i \\
 & 1 & 0.582643803-0.281298113 i & 0.582643867-0.281297980 i &
   0.582643988-0.281298073 i & 0.582642026-0.281290444 i \\
  & 2 & 0.551684901-0.479092751 i & 0.551685313-0.479090887 i &
   0.551686148-0.479091046 i & 0.551593902-0.479046852 i \\
 & 3 & 0.511961911-0.690337096 i & 0.512033346-0.690300137 i &
   0.511933962-0.690327581 i & 0.511099447-0.69046816 i \\
 & 4 & 0.470174006-0.915649393 i & 0.470522642-0.915732579 i &
   0.470104354-0.915739619 i & 0.466858911-0.917951148 i \\
  & 5 & 0.431386479-1.15215136 i & 0.432090440-1.15222942 i &
   0.431350740-1.15245777 i & 0.424319802-1.16246037 i \\
 4 & 0 & 0.809178378-0.094163961 i & 0.809178378-0.0941639610 i &
   0.809178378-0.0941639610 i & 0.809178349-0.0941640697 i \\
  & 1 & 0.796631532-0.284334349 i & 0.796631531-0.284334349 i &
   0.796631531-0.284334348 i & 0.796631254-0.284334155 i \\
  & 2 & 0.772709533-0.479908175 i & 0.772709546-0.479908196 i &
   0.772709526-0.479908164 i & 0.772695364-0.479899647 i \\
  & 3 & 0.73983673-0.683924319 i & 0.739836320-0.683925183 i &
   0.739836665-0.683924261 i & 0.739665055-0.683901582 i \\
  & 4 & 0.701515509-0.898238972 i & 0.701514029-0.898239961 i &
   0.701514717-0.898238277 i & 0.700640747-0.898460471 i \\ 
\end{tabular} 
\end{ruledtabular}
 \end{table}
\endgroup

\begingroup\squeezetable
\begin{table}
 \caption{\label{tabb5}   The quasinormal modes of the even-parity 
 gravitational perturbations of the Schwarzschild black hole. Numerical results are 
 taken from Ref.~\cite{phase1}.} 
\begin{ruledtabular}
\begin{tabular}{cccccc}
$ l$ & $n$ & \text{Numerical value of $\omega$} & Zerilli potential $P_{6}^{6}$  & Zerilli potential $P_{7}^{6}$ &
   {\rm Sixth order WKB}  \\ \hline
 2 & 0 & 0.373671684-0.088962315 i & 0.373671632-0.0889623659 i &
   0.373671627-0.0889624367 i & 0.373707325-0.0889233432 i \\
  & 1 & 0.346710997-0.273914875 i & 0.346675500-0.273936180 i &
   0.346710669-0.273910076 i & 0.346715435-0.273875995 i \\
  & 2 & 0.301053455-0.478276983 i & 0.301080664-0.478208204 i &
   0.301133004-0.478206943 i & 0.300049631-0.478828264 i \\
  & 3 & 0.251504962-0.705148202 i & 0.252152974-0.704767125 i &
   0.250186809-0.703956848 i & 0.245512646-0.711587168 i \\
 3 & 0 & 0.599443288-0.092703048 i & 0.599443288-0.0927030481 i &
   0.599443288-0.0927030481 i & 0.599443373-0.0927027654 i \\
  & 1 & 0.582643803-0.281298113 i & 0.582643798-0.281298116 i &
   0.582643797-0.281298115 i & 0.582642128-0.281292695 i \\
  & 2 & 0.551684901-0.479092751 i & 0.551685635-0.479091963 i &
   0.551685178-0.479091911 i & 0.551595711-0.479055726 i \\
  & 3 & 0.511961911-0.690337096 i & 0.511964029-0.690332484 i &
   0.511965550-0.690331591 i & 0.511107845-0.690490597 i \\
  & 4 & 0.470174006-0.915649393 i & 0.470171037-0.915616065 i &
   0.470157447-0.915641228 i & 0.466882696-0.917994938 i \\
  & 5 & 0.431386479-1.15215136 i & 0.431254524-1.15207234 i &
   0.431276665-1.15224185 i & 0.424371238-1.16253344 i \\
 4 & 0 & 0.809178378-0.094163961 i & 0.809178378-0.0941639610 i &
   0.809178378-0.0941639610 i & 0.809178349-0.0941640733 i \\
  & 1 & 0.796631532-0.284334349 i & 0.796631532-0.284334349 i &
   0.796631532-0.284334349 i & 0.796631252-0.284334186 i \\
  & 2 & 0.772709533-0.479908175 i & 0.772709540-0.479908166 i &
   0.772709544-0.479908170 i & 0.772695376-0.479899772 i \\
  & 3 & 0.73983673-0.683924319 i & 0.739836652-0.683924275 i &
   0.739836754-0.683924252 i & 0.739665134-0.683901916 i \\
  & 4 & 0.701515509-0.898238972 i & 0.701515239-0.898238200 i &
   0.701515431-0.898238235 i & 0.700641016-0.898461166 i \\
\end{tabular}
\end{ruledtabular}
 \end{table}
\endgroup

\begingroup\squeezetable
\begin{table}
 \caption{\label{tabb5a} Deviations of the real and imaginary  part 
 of the quasinormal frequencies of the even-parity gravitational 
 perturbations for  the Schwarzschild black hole from the accurate 
 numerical results. The Pad\'e approximation always gives better results
 that the 6-th order WKB.}
\begin{ruledtabular}
\begin{tabular}{cccccccc}
&  & $P_{6}^{6}$ & $P_{6}^{6}$ & $P_{7}^{6} $ & $P_{7}^{6}$ &  {\rm Sixth order WKB}  &  {\rm Sixth order WKB} \\
 $l$ & $n$ & $\Delta^{(r)}$   $(\%)$  & $\Delta^{(i)}$ $(\%)$ & $\Delta^{(r)}$  $(\%)$ & $\Delta^{(i)}$  $(\%)$ & $\Delta^{(r)}$  $(\%)$ & $\Delta^{(i)}$ $(\%)$ \\ \hline 
 $2 $ & $ 0 $ & $ 1.4\times 10^{-5} $ & $ -5.7\times 10^{-5} $ & $ 1.5\times 10^{-5} $ & $ -1.4\times
   10^{-4} $ & $ 1.4\times 10^{-2} $ & $ 8.\times 10^{-2}$ \\
$  $ & $ 1 $ & $ 1.\times 10^{-2} $ & $ -7.8\times 10^{-3} $ & $ 9.4\times 10^{-5} $ & $
   1.8\times 10^{-3} $ & $ 1.2\times 10^{-1} $ & $ 1.6\times 10^{-1} $\\
$  $ & $ 2 $ & $ -9.\times 10^{-3} $ & $ 1.4\times 10^{-2} $ & $ -2.6\times 10^{-2} $ & $
   1.5\times 10^{-2} $ & $ 8.4\times 10^{-1} $ & $ 1.5\times 10^{-1} $\\
$  $ & $ 3 $ & $ -2.6\times 10^{-1} $ & $ 5.4\times 10^{-2} $ & $ 5.2\times 10^{-1} $ & $
   1.7\times 10^{-1} $ & $ 3.9 $ & $ -6.3\times 10^{-1} $\\
$ 3 $ & $ 0 $ & $ -6.9\times 10^{-8} $ & $ -6.5\times 10^{-8} $ & $ -6.6\times 10^{-8} $ & $ -1.3\times
   10^{-7} $ & $ -1.4\times 10^{-5} $ & $ 5.7\times 10^{-4}$ \\
$  $ & $ 1 $ & $ 9.3\times 10^{-7} $ & $ -1.1\times 10^{-6} $ & $ 1.\times 10^{-6} $ & $
   -7.8\times 10^{-7} $ & $ 3.\times 10^{-4} $ & $ 2.7\times 10^{-3}$ \\
$  $ & $ 2 $ & $ -1.3\times 10^{-4} $ & $ 1.6\times 10^{-4} $ & $ -5.\times 10^{-5} $ & $
   1.8\times 10^{-4} $ & $ 1.6\times 10^{-2} $ & $ 9.6\times 10^{-3} $ \\
$  $ & $ 3 $ & $ -4.1\times 10^{-4} $ & $ 6.7\times 10^{-4} $ & $ -7.1\times 10^{-4} $ & $
   8.\times 10^{-4} $ & $ 1.7\times 10^{-1} $ & $ -1.9\times 10^{-2} $ \\
$  $ & $ 4 $ & $ 6.3\times 10^{-4} $ & $ 3.6\times 10^{-3} $ & $ 3.5\times 10^{-3} $ & $
   8.9\times 10^{-4} $ & $ 7.1\times 10^{-1} $ & $ -2.5\times 10^{-1} $ \\
$  $ & $ 5 $ & $ 3.1\times 10^{-2} $ & $ 6.9\times 10^{-3} $ & $ 2.5\times 10^{-2} $ & $
   -7.9\times 10^{-3} $ & $ 1.6 $ & $ -8.9\times 10^{-1} $  \\
$ 4 $ & $ 0 $ & $ 5.7\times 10^{-8} $ & $ 1.8\times 10^{-8} $ & $ 5.8\times 10^{-8} $ & $ 1.4\times
   10^{-8} $ & $ 3.6\times 10^{-6} $ & $ -1.2\times 10^{-4} $ \\
$  $ & $ 1 $ & $ 4.1\times 10^{-8} $ & $ -1.1\times 10^{-7} $ & $ -1.5\times 10^{-8} $ & $
   -9.8\times 10^{-8} $ & $ 3.5\times 10^{-5} $ & $ 6.8\times 10^{-5}  $\\
$  $ & $ 2 $ & $ -9.\times 10^{-7} $ & $ 1.8\times 10^{-6} $ & $ -1.4\times 10^{-6} $ & $
   1.1\times 10^{-6} $ & $ 1.8\times 10^{-3} $ & $ 1.8\times 10^{-3}  $ \\
$  $ & $ 3 $ & $ 1.\times 10^{-5} $ & $ 6.4\times 10^{-6} $ & $ -3.2\times 10^{-6} $ & $
   9.8\times 10^{-6} $ & $ 2.3\times 10^{-2} $ & $ 3.3\times 10^{-3} $ \\
$  $ & $ 4 $ & $ 3.9\times 10^{-5} $ & $ 8.6\times 10^{-5} $ & $ 1.1\times 10^{-5} $ & $
   8.2\times 10^{-5} $ & $ 1.2\times 10^{-1} $ & $ -2.5\times 10^{-2} $ \\
\end{tabular}
\end{ruledtabular}
 \end{table}
\endgroup

\begingroup\squeezetable
\begin{table}
\caption{\label{tabb6}  The quasinormal $l=2$ modes of the odd-parity electromagnetic 
perturbations of the Reissner-Nordstr\"om  black hole for several exemplary values of $q=|e|/M.$  
The numerical results are 
 taken from Ref.~\cite{AnderssonRN1}}
\begin{ruledtabular}
\begin{tabular}{ccccc}
Method & $q$  & $n=0$ & $n=1$ & $n=2$ \\ \hline
 numerical  & 0.1 & 0.45892629-0.095097192 i & 0.4379399-0.29096894 i & \
0.40269292-0.50195535 i \\
 $P_{6}^{6}$ &  & 0.45892629-0.095097193 i & 0.43793984-0.29096895 i \
& 0.40269555-0.50195231 i \\
 $P_{7}^{6}$ &  & 0.45892629-0.095097192 i & 0.43793988-0.29096896 i \
& 0.40269189-0.50195635 i \\ \hline
 numerical  & 0.2 & 0.46296518-0.095373436 i & 0.4421817-0.29173823 i & \
0.4072649-0.5030451 i \\
 $P_{6}^{6}$ &  & 0.46296518-0.095373436 i & 0.44218165-0.29173825 i \
& 0.40726746-0.50304215 i \\
 $P_{7}^{6}$ &  & 0.46296518-0.095373436 i & 0.44218168-0.29173825 i \
& 0.40726409-0.50304616 i \\ \hline
 numerical  & 0.3 & 0.46986497-0.09582639 i & 0.44942963-0.29299464 i & \
0.41507862-0.50480957 i \\
 $P_{6}^{6}$ &  & 0.46986497-0.095826393 i & 0.44942959-0.29299467 i \
& 0.41508170-0.50480617 i \\
 $P_{7}^{6}$ &  & 0.46986497-0.095826390 i & 0.44942958-0.29299467 i \
& 0.41507668-0.50481216 i \\ \hline
 numerical  & 0.4 & 0.47992596-0.096442085 i & 0.46000231-0.29469011 i & \
0.42648154-0.50715223 i \\
 $P_{6}^{6}$ &  & 0.47992596-0.096442086 i & 0.46000227-0.29469013 i \
& 0.42648483-0.50714702 i \\
 $P_{7}^{6}$ &  & 0.47992596-0.096442085 i & 0.46000230-0.29469013 i \
& 0.42647371-0.50715168 i \\ \hline
 numerical  & 0.5 & 0.49367474-0.09719237 i & 0.4744594-0.29672916 i & \
0.44208527-0.50988382 i \\
 $P_{6}^{6}$ &  & 0.49367474-0.097192370 i & 0.47445937-0.29672916 i \
& 0.44208770-0.50987907 i \\
 $P_{7}^{6}$ &  & 0.49367474-0.097192370 i & 0.47445941-0.29672917 i \
& 0.44208007-0.50988121 i \\ \hline
 numerical  & 0.6 & 0.51201092-0.098016629 i & 0.49375723-0.29891032 i & \
0.46293471-0.51261437 i \\
 $P_{6}^{6}$ &  & 0.51201092-0.098016630 i & 0.49375723-0.29891041 i \
& 0.46293268-0.51261303 i \\
 $P_{7}^{6}$ &  & 0.51201092-0.098016630 i & 0.49375717-0.29891035 i \
& 0.46293338-0.51261246 i \\ \hline 
 numerical  & 0.7 & 0.53650775-0.098771413 i & 0.5195641-0.30076675 i & \
0.4908422-0.51446414 i l \\
 $P_{6}^{6}$ &  & 0.53650775-0.098771415 i & 0.51956416-0.30076686 i \
& 0.49084282-0.51446162 i \\
 $P_{7}^{6}$ &  & 0.53650775-0.098771413 i & 0.51956400-0.30076675 i \
& 0.49084016-0.51446349 i \\ \hline
 numerical  & 0.8 & 0.57013023-0.099069062 i & 0.5549928-0.30105923 i & \
0.52912062-0.51315436 i \\
 $P_{6}^{6}$ &  & 0.57013023-0.099069062 i & 0.55499292-0.30105933 i \
& 0.52912089-0.51315308 i \\
 $P_{7}^{6}$ &  & 0.57013023-0.099069062 i & 0.55499274-0.30105923 i \
& 0.52911993-0.51315423 i \\ \hline
 numerical  & 0.9 & 0.61939761-0.09758287 i & 0.6066257-0.29560854 i & \
0.58421715-0.50115933 i \\
 $P_{6}^{6}$ &  & 0.61939761-0.097582871 i & 0.60662569-0.29560853 i \
& 0.58421701-0.50115896 i \\
 $P_{7}^{6}$ &  & 0.61939761-0.097582870 i & 0.60662570-0.29560853 i \
& 0.58421705-0.50115890 i \\ \hline
 numerical  & 0.99 & 0.69275202-0.088642205 i & 0.6786593-0.26750387 i & \
0.65095306-0.45118039 i \\
 $P_{6}^{6}$ &  & 0.69275202-0.088642191 i & 0.67865945-0.26750386 i \
& 0.65095149-0.45118011 i \\
 $P_{7}^{6}$ &  & 0.69275202-0.088642204 i & 0.67865927-0.26750401 i \
& 0.65095227-0.45118114 i \\ 
\end{tabular}
\end{ruledtabular}
 \end{table}
\endgroup

\begingroup\squeezetable
\begin{table}
 \caption{\label{tabb7}   The quasinormal $l=2$ modes of the odd-parity gravitational perturbations 
 of the Reissner-Nordstr\"om  black hole for several exemplary values of $q=|e|/M.$  
The numerical results are  taken from Ref.~\cite{AnderssonRN1}}
\begin{ruledtabular}
\begin{tabular}{ccccc}
Method & $q$  & $n=0$ & $n=1$ & $n=2$ \\ \hline
 numerical  & 0.1 & 0.37393238-0.088990533 i & 0.34698103-0.27399495 i & \
0.3013348-0.47839381 i \\
 $P_{6}^{6}$ &  & 0.37393222-0.088998960 i & 0.34655495-0.27401278 i \
& 0.29943057-0.47869569 i \\
 $P_{7}^{6}$ &  & 0.37393371-0.088994903 i & 0.34700547-0.27385072 i \
& 0.30202310-0.47667901 i \\ \hline
 numerical  & 0.2 & 0.37474443-0.089074786 i & 0.34782603-0.27423284 i & \
0.30222215-0.47873633 i \\
 $P_{6}^{6}$ &  & 0.37474440-0.089083239 i & 0.34740014-0.27426511 i \
& 0.30027463-0.47911441 i \\
 $P_{7}^{6}$ &  & 0.37474583-0.089079319 i & 0.34785085-0.27408448 i \
& 0.30286554-0.47700100 i \\ \hline
 numerical  & 0.3 & 0.37619678-0.089212946 i & 0.34935008-0.27461849 i & \
0.30384543-0.47927509 i \\
 $P_{6}^{6}$ &  & 0.37619695-0.089221471 i & 0.34892699-0.27467057 i \
& 0.30184885-0.47976635 i \\
 $P_{7}^{6}$ &  & 0.37619829-0.089217694 i & 0.34937482-0.27446442 i \
& 0.30441295-0.47751621 i \\ \hline
 numerical  & 0.4 & 0.37843689-0.089398114 i & 0.35172753-0.2751243 i & \
0.30642431-0.47994045 i \\
 $P_{6}^{6}$ &  & 0.37843726-0.089406811 i & 0.35131155-0.27519530 i \
& 0.30440527-0.48055153 i \\
 $P_{7}^{6}$ &  & 0.37843856-0.089403029 i & 0.35175035-0.27496564 i \
& 0.30688980-0.47817569 i \\ \hline
 numerical  & 0.5 & 0.38167715-0.089612379 i & 0.35521299-0.27568438 i & \
0.31028277-0.48058204 i \\
 $P_{6}^{6}$ &  & 0.38167772-0.089621446 i & 0.35480784-0.27576354 i \
& 0.30831023-0.48125731 i \\
 $P_{7}^{6}$ &  & 0.38167903-0.089617226 i & 0.35522981-0.27552809 i \
& 0.31064235-0.47886552 i \\ \hline
 numerical  & 0.6 & 0.38621747-0.089813675 i & 0.36017029-0.27615005 i & \
0.31588482-0.48087942 i \\
 $P_{6}^{6}$ &  & 0.38621829-0.089823511 i & 0.35977578-0.27620272 i \
& 0.31409566-0.48143936 i \\
 $P_{7}^{6}$ &  & 0.38621953-0.089817849 i & 0.36017500-0.27601248 i \
& 0.31619103-0.47931687 i \\ \hline
 numerical  & 0.7 & 0.39249849-0.089904426 i & 0.36713339-0.27618218 i & \
0.32389361-0.48012132 i \\
 $P_{6}^{6}$ &  & 0.39250093-0.089915592 i & 0.36681075-0.27606507 i \
& 0.32266295-0.48012476 i \\
 $P_{7}^{6}$ &  & 0.39250008-0.089906645 i & 0.36712177-0.27608144 i \
& 0.32427673-0.47886675 i \\ \hline
 numerical  & 0.8 & 0.40121719-0.08964323 i & 0.37690401-0.27494337 i & \
0.33517151-0.47656389 i \\
 $P_{6}^{6}$ &  & 0.40122292-0.089647871 i & 0.37692966-0.27485827 i \
& 0.33521872-0.47580206 i \\
 $P_{7}^{6}$ &  & 0.40121726-0.089643579 i & 0.37693533-0.27482471 i \
& 0.33570503-0.47578542 i \\ \hline
 numerical  & 0.9 & 0.4135705-0.088333012 i & 0.39045691-0.27001504 i & \
0.34959435-0.46521762 i \\
 $P_{6}^{6}$ &  & 0.41357316-0.088335816 i & 0.39049403-0.26997448 i \
& 0.35028615-0.46480977 i \\
 $P_{7}^{6}$ &  & 0.41357020-0.088332922 i & 0.39059923-0.26997219 i \
& 0.35064801-0.46456627 i \\ \hline
 numerical  & 0.99 & 0.42929679-0.084265944 i & 0.40352005-0.25701425 i & \
0.35391255-0.44357761 i \\
 $P_{6}^{6}$ &  & 0.42929725-0.084268033 i & 0.40356088-0.25700759 i \
& 0.35441725-0.44384073 i \\
 $P_{7}^{6}$ &  & 0.42929632-0.084265372 i & 0.40355372-0.25700435 i \
& 0.35539583-0.44397284 i \\ 
\end{tabular}
\end{ruledtabular}
 \end{table}
\endgroup

\begingroup\squeezetable
\begin{table}
 \caption{\label{tabb8} The quasinormal  modes of the odd-parity gravitational perturbations of the extreme $(|e|=M)$  Reissner-Nordstr\"om  black hole.
The numerical results are  taken from Ref.~\cite{Onozawa1}. To conform with~\cite{Onozawa1} we have rounded our results appropriately.}
\begin{ruledtabular}
\begin{tabular}{ccccc}
 & Method  & $n=0$ & $n=1$ & $n=2$ \\ \hline
 $l=2$ & numerical & 0.43134-0.083460 i & 0.40452-0.25498 i & 0.35340-0.44137 i \\
 & $P_{6}^{6} $ & 0.43134-0.083462 i & 0.40457-0.25498  i & 0.35412-0.44167 i \\
 & $P_{7}^{6}$ & 0.43134-0.083460 i & 0.40459-0.25499 i & 0.35515-0.44191 i \\
 $l=3$ & numerical & 0.70430-0.085973 i & 0.68804-0.25992 i & 0.65624-0.44007 i \\
 & $P_{6}^{6}$ & 0.70430-0.085973 i & 0.68804-0.25992 i & 0.65624-0.44007 i \\
 & $P_{7}^{6}$ & 0.70430-0.085973 i & 0.68804-0.25992 i & 0.65624-0.44007 i \\
$l=4$ & numerical & 0.96576-0.087001 i & 0.95381-0.26212 i & 0.93020-0.44064 i \\
 & $P_{6}^{6}$ & 0.96576-0.087001 i & 0.95381-0.26212 i & 0.93020-0.44064 i \\
 & $P_{7}^{6}$ & 0.96576-0.087001 i & 0.95381-0.26212 i & 0.93020-0.44064 i \\
\end{tabular}
\end{ruledtabular}
 \end{table}

 \begingroup\squeezetable
\begin{table}
 \caption{\label{tabb9}   Deviations of the real and imaginary  part of the Pad\'e approximants, $P_{6}^{6},$ from the accurate numerical results.
 The $(l=2)$ quasinormal frequencies of the odd-parity gravitational perturbations of the Reissner-Nordstr\"om black holes are given in Table~\ref{tabb6}.}
\begin{ruledtabular}
 \begin{tabular}{ccccccc}
 & $\Delta^{(r)}$ $(\%)$ & $\Delta^{(i)}$ $(\%)$ & $\Delta^{(r)}$ $(\%)$ & $\Delta^{(i)}$ $(\%)$ & $\Delta^{(r)}$ $(\%)$ & $\Delta^{(i)}$ $(\%)$ \\
 $q$        &      $n=0$           &    $n=0$                  & $       n=1         $ & $   n=1                  $ & $   n=2                $ & $    n=2 $       \\ \hline
 $ 0.1  $ & $   4.\times 10^{-5} $ & $ -9.5\times 10^{-3} $ & $ 1.2\times 10^{-1} $ & $ -6.5\times 10^{-3} $ & $ 6.3\times 10^{-1} $ & $ -6.3\times 10^{-2}$ \\
 $ 0.2  $ & $ 6.4\times 10^{-6} $ & $ -9.5\times 10^{-3} $ & $ 1.2\times 10^{-1} $ & $ -1.2\times 10^{-2} $ & $ 6.4\times 10^{-1} $ & $ -7.9\times 10^{-2}$ \\
 $ 0.3  $ & $ -4.3\times 10^{-5} $ & $ -9.6\times 10^{-3} $ & $ 1.2\times 10^{-1} $ & $ -1.9\times 10^{-2} $ & $ 6.6\times 10^{-1} $ & $ -1.\times 10^{-1} $\\
 $ 0.4  $ & $ -9.9\times 10^{-5} $ & $ -9.7\times 10^{-3} $ & $ 1.2\times 10^{-1} $ & $ -2.6\times 10^{-2} $ & $ 6.6\times 10^{-1} $ & $ -1.3\times 10^{-1} $\\
 $ 0.5  $ & $ -1.5\times 10^{-4} $ & $ -1.\times 10^{-2} $ & $ 1.1\times 10^{-1} $ & $ -2.9\times 10^{-2} $ & $ 6.4\times 10^{-1} $ & $ -1.4\times 10^{-1} $\\
 $ 0.6  $ & $ -2.1\times 10^{-4} $ & $ -1.1\times 10^{-2} $ & $ 1.1\times 10^{-1} $ & $ -1.9\times 10^{-2} $ & $ 5.7\times 10^{-1} $ & $ -1.2\times 10^{-1}$ \\
 $ 0.7  $ & $ -6.2\times 10^{-4} $ & $ -1.2\times 10^{-2} $ & $ 8.8\times 10^{-2} $ & $ 4.2\times 10^{-2} $ & $ 3.8\times 10^{-1} $ & $ -7.2\times 10^{-4} $\\
 $ 0.8  $ & $ -1.4\times 10^{-3} $ & $ -5.2\times 10^{-3} $ & $ -6.8\times 10^{-3} $ & $ 3.1\times 10^{-2} $ & $ -1.4\times 10^{-2} $ & $ 1.6\times 10^{-1} $\\
 $ 0.9  $ & $ -6.4\times 10^{-4} $ & $ -3.2\times 10^{-3} $ & $ -9.5\times 10^{-3} $ & $ 1.5\times 10^{-2} $ & $ -2.\times 10^{-1} $ & $ 8.8\times 10^{-2} $\\
 $ 0.99 $ & $ -1.1\times 10^{-4} $ & $ -2.5\times 10^{-3} $ & $ -1.\times 10^{-2} $ & $ 2.6\times 10^{-3} $ & $ -1.4\times 10^{-1} $ & $ -5.9\times 10^{-2}$ \\
\end{tabular}
\end{ruledtabular}
 \end{table}


 \begingroup\squeezetable
\begin{table}
 \caption{\label{tabb10}  Deviations of the real and imaginary  part of the Pad\'e approximants, $P_{7}^{6},$ from the accurate numerical results.
 The $(l=2)$ quasinormal frequencies of the odd-parity gravitational perturbations of the Reissner-Nordstr\"om black holes are given in Table~\ref{tabb7}.}
\begin{ruledtabular}
\begin{tabular}{ccccccc}
& $\Delta^{(r)}$ $(\%)$ & $\Delta^{(i)}$ $(\%)$ & $\Delta^{(r)}$ $(\%)$ & $\Delta^{(i)}$ $(\%)$ & $\Delta^{(r)}$ $(\%)$ & $\Delta^{(i)}$ $(\%)$ \\
 $q$        &      $n=0$           &    $n=0$                  & $       n=1         $ & $   n=1                  $ & $   n=2                $ & $    n=2 $       \\ \hline
$ 0.1 $ & $ -3.6\times 10^{-4} $ & $ -4.9\times 10^{-3} $ & $ -7.\times 10^{-3} $ & $ 5.3\times 10^{-2} $ & $ -2.3\times 10^{-1} $ & $ 3.6\times 10^{-1} $\\
$ 0.2 $ & $ -3.7\times 10^{-4} $ & $ -5.1\times 10^{-3} $ & $ -7.1\times 10^{-3} $ & $ 5.4\times 10^{-2} $ & $ -2.1\times 10^{-1} $ & $ 3.6\times 10^{-1} $\\
$ 0.3 $ & $ -4.\times 10^{-4} $ & $ -5.3\times 10^{-3} $ & $ -7.1\times 10^{-3} $ & $ 5.6\times 10^{-2} $ & $ -1.9\times 10^{-1} $ & $ 3.7\times 10^{-1}  $\\
$ 0.4 $ & $ -4.4\times 10^{-4} $ & $ -5.5\times 10^{-3} $ & $ -6.5\times 10^{-3} $ & $ 5.8\times 10^{-2} $ & $ -1.5\times 10^{-1} $ & $ 3.7\times 10^{-1} $ \\
$ 0.5 $ & $ -4.9\times 10^{-4} $ & $ -5.4\times 10^{-3} $ & $ -4.7\times 10^{-3} $ & $ 5.7\times 10^{-2} $ & $ -1.2\times 10^{-1} $ & $ 3.6\times 10^{-1} $ \\
$ 0.6 $ & $ -5.3\times 10^{-4} $ & $ -4.6\times 10^{-3} $ & $ -1.3\times 10^{-3} $ & $ 5.\times 10^{-2} $ & $ -9.7\times  10^{-2} $ & $ 3.2\times 10^{-1} $ \\
$ 0.7 $ & $ -4.\times 10^{-4} $ & $ -2.5\times 10^{-3} $ & $ 3.2\times 10^{-3} $ & $ 3.6\times 10^{-2} $ & $ -1.2\times 10^{-1} $ & $ 2.6\times 10^{-1} $\\
$ 0.8 $ & $ -1.9\times 10^{-5} $ & $ -3.9\times 10^{-4} $ & $ -8.3\times 10^{-3} $ & $ 4.3\times 10^{-2} $ & $ -1.6\times 10^{-1} $ & $ 1.6\times 10^{-1} $ \\
$ 0.9 $ & $ 7.3\times 10^{-5} $ & $ 1.\times 10^{-4} $ & $ -3.6\times 10^{-2} $ & $ 1.6\times 10^{-2} $ & $ -3.\times 10^{-1} $ & $ 1.4\times 10^{-1} $\\
$ 0.99 $ & $ 1.1\times 10^{-4} $ & $ 6.8\times 10^{-4} $ & $ -8.3\times 10^{-3} $ & $ 3.9\times 10^{-3} $ & $ -4.2\times 10^{-1} $ & $ -8.9\times 10^{-2} $\\
\end{tabular}
\end{ruledtabular}
 \end{table}
 
 \section{Concluding remarks\label{last}} 
We have used the the thirteenth order WKB method and the Pad\'e transforms to calculate the quasinormal
modes of the Schwarzschild and Reissner-Nordstr\"om black holes and demonstrated that our results are 
very close to the accurate numerical calculations. The method can be modified to allow for more complicated potentials,
with the function $V$ depending on $\omega.$
Our general formulas can be applied straightforwardly to any black hole potential given by Eq.~(\ref{narrow}) and the only 
limit of the  calculations is their scale.  For example, the quasinormal modes of the $d$-dimensional
Schwarzschild-Tangherlini black holes~\cite{jose4}, the asymptotically (anti)-de Sitter~\cite{jose1,jose2,jose3} black holes  
and many others can be calculated without necessity to 
change the codes.
It should be emphasized that some potentials may  not be  strictly positive and their complexity grows with the dimension.

Preliminary calculations and first comparisons with the known results look promising. 
Indeed, our calculations of the fundamental tensor gravitational quasinormal modes of the higher dimensional Schwarzschil-Tangherlini 
black holes $(5\leq d \leq 11)$ give exactly the same results as those presented in Refs.~\cite{Rostworowski,Zhidenko}.
In the $d=11$-dimensional case 
for $2\leq l \leq 11$ and the lowest overtones the Pad\'e approximants yield  good results. For example, for $l=2$ and $n=0,1,2$
the Pad\'e approximants $P_{6}^{6}$ (rounded to four decimal places) give, respectively,  $\omega =4.3920 - 1.0577 i,$
$\omega = 3.3393 - 3.0283 i$ and $\omega = 1.8026 - 3.6527 i,$
which is identical $(l=0,n=0)$ or close to the frequencies calculated  by Rostworowski~\cite{Rostworowski} and
the accuracy of our calculations rapidly grows with $l.$ The behavior of the low-lying modes may be contrasted
with the WKB approximation. For example for $l=2$ one has $\omega = 4.4007 - 1.0601 i$, $\omega =3.1165 - 3.7864 i $ and $\omega = 0.5276 - 5.6632 i $. 
The real part of the frequency of the mode $l=2$ and $n=2$ is wrong at any order of the WKB.
Specifically, within the sixth order WKB approximation, it is over three times smaller than the accurate 
numerical value calculated by Rostworowski.  On the other hand, however, the approximation works better for larger $l,$
as expected. This comparison indicates that one should be cautious with any approximation based on the WKB method even for the 
low overtones ($n = 2$) of the low-lying modes $l=2$ and $l=3$. 
On the other hand, our results demonstrate  the usefulness of the Pad\'e approximation and suggest that there is still a room for improvements and new ideas.
In our personal view the method advocated by Matiukhin and Gal'tsov~\cite{galtsov} and its possible generalizations, although technically very hard, 
look promising. We also observe that the calculated functions  $\Lambda_{k}$ can be applied in the analysis of the potential barrier tunneling.
These problems are actively investigated and the results  will be presented elsewhere.
Finally, observe that in the calculations of the quasinormal modes one often 
has to choose between the generality and simplicity of the approach on 
the one hand and the great accuracy on the other, and, consequently, the quality of the approximation
has  be judged not only by comparison with the exact numerical results but also with the competing analytic or semianalytic approaches.

\begin{acknowledgments}
Discussions with Waldek Berej are gratefully acknowledged.
J.M. was partially supported by the Polish National Science Centre grant no. DEC-2014/15/B/ST2/00089.
\end{acknowledgments}

\clearpage

\end{document}